# Production of neutron-rich heavy residues and the freeze-out temperature in the fragmentation of relativistic $^{238}$U projectiles determined by the isospin thermometer[*]

K.-H. Schmidt[†], M. V. Ricciardi, A. S. Botvina[‡], T. Enqvist[§]

*GSI, Planckstr. 1, 64291 Darmstadt, Germany*

**Abstract:** Isotope yields of heavy residues produced in collisions of $^{238}$U with lead at 1·$A$ GeV show indications for a simultaneous break-up process. From the average $N$-over-$Z$ ratio of the final residues up to $Z = 70$, the average limiting temperature of the break-up configuration at freeze out was determined to $T \approx 5$ MeV using the isospin-thermometer method. Consequences for the understanding of other phenomena in highly excited nuclear systems are discussed.

## 1. Introduction

The decay of highly excited nuclei takes place in many processes, such as fission, spallation, fragmentation and others, which have large practical importance. As was shown long ago within statistical approaches (see e.g. [1]), the compound-nucleus mechanism dominating at low excitation energies gradually changes into simultaneous decay into many fragments (multifragmentation) with increasing energy. The multifragmentation process can be associated with the liquid-gas-type phase transition of nuclear matter [1], and many signatures of this phenomenon have been accumulated (see e.g. [2, 3]). These are the total disintegration of nuclei into intermediate-mass and light fragments, taking place in a short time and described by statistical models, scaling laws for the size of the fragments, critical phenomena manifested as specific correlations between fragments of different size, the caloric curve with nearly constant temperature in the wide excitation-energy range where the nuclei totally disintegrate, the large magnitude of energy fluctuations in fragment partitions, signatures for the formation of equal-size fragments in the spinodal region and others. Recently, encouraging results were obtained in searching for the critical temperature of this phase transition [4, 5]. It has also been discussed to measure the ratio of protons and neutrons in the fragmentation of very asymmetric nuclear matter in order to detect chemical instabilities which lead to a separation between the proton and the neutron phase by diffuse spinodal and isospin fractionation [6, 7]. All these signatures are intimately related to *multifragmentation*, i.e. the formation of relatively light clusters in a simultaneous break-up.

---

[*] This work forms part of the PhD thesis of M. V. Ricciardi.
[†] Corresponding author, e-mail: K.H.Schmidt@gsi.de, WEB: www-wnt.gsi.de/kschmidt
[‡] Permanent address: Inst. for Nuclear Research, Russian Academy of Sciences, 117312 Moscow, Russia
[§] Present address: University of Jyväskylä, 40351 Jyväskylä, Finland

While the activities have concentrated on the investigation of light clusters so far, the present work intends to extract complementary information by studying the formation of heavy residues. Heavy residues are clearly associated with the liquid phase. Heavy clusters are sensitive to the temperature of a possible phase transition and their sizes are directly connected with the thermal excitation energy of the system. In this way we exclude from our analysis a contribution from the gaseous phase, which may extend to higher temperatures and which is difficult to identify on the background of preequilibrium processes. We take into consideration some experimental data on heavy-fragment production, which earlier was considered only as a result of a spallation-type process. In particular, we investigate manifestations of the liquid-gas-type phase transition by analysing the formation of heavy residues in the fragmentation of $^{238}$U in a lead target, measured in a previous experiment [8], with the isospin thermometer. With this method, the mean "isospin" (Z-N)/2, or as a related quantity the mean N-over-Z ratio of the final fragments is used in combination with statistical-model codes in order to deduce the freeze-out temperature after break up.

We will describe the technical aspects of the experiment (section 2), discuss the sensitivity of the N-over-Z ratio of the fragments to a simultaneous-emission phase (section 3.1), present the idea of the isospin thermometer to deduce the temperature at the beginning of the sequential decay (section 3.2.1) and analyse the data with three different approaches, a schematic model (sections 3.2.2), a three-stage nuclear-reaction model (section 3.3), and the statistical multi-fragmentation model (section 3.4). Finally, we propose a possible scenario of mid-peripheral high-energy nucleus-nucleus collisions (section 3.5) and discuss the consequences for the understanding of related phenomena (sections 3.6 and 3.7).

## 2. Experiment and Results

### 2.1. The fragment separator and the detection system

The experiment [8] was performed with the high-resolution, magnetic forward spectrometer, the fragment separator (FRS) [9], at GSI. The FRS is a two-stage spectrometer with a dispersive intermediate image plane and an achromatic final image plane. Its angular acceptance is about 15 mrad around the beam axis, and its momentum acceptance amounts to about 3%.

The $^{238}$U primary beam, delivered from the SIS synchrotron at the energy of 1·A GeV impinged on a lead target of 50.5 mg/cm$^2$ placed at the entrance of the FRS in the experiment considered [8]. A secondary-electron transmission monitor (SEETRAM) [10] was constantly used to measure the beam intensity of up to $10^7$ ions/s. The detection equipment (see figure 1) consisted mainly of two scintillation detectors, placed at the intermediate image plane and at the exit, and of an ionisation chamber, placed behind the FRS. The ionisation chamber recorded the energy loss $\Delta E$ of the produced ions, giving a measurement of its nuclear charge, Z. The two scintillation detectors were used to detect the horizontal positions as well as the time-of-flight between the mid-plane and the exit. The information on the positions at mid and final plane gave a measurement of the magnetic rigidity ($B\rho$) of the ions passing through the FRS.

Among all the fragments produced in the interaction of the beam with the target, only those with a certain magnetic rigidity, selected by tuning the fields of the first and second dipole, could pass the spectrometer. Due to the energy loss in the first scintillator, all the fragments reduced their energy and only those with some selected *Z* range reached the exit and passed in a definite position at the exit. Thus, the magnetic field in the second half of the spectrometer decided the range of the transmitted elements. Because of the limited acceptance in magnetic rigidity of the FRS, a combination of several *Bρ* settings was needed in order to cover the full range of *A/Z* and of velocity.

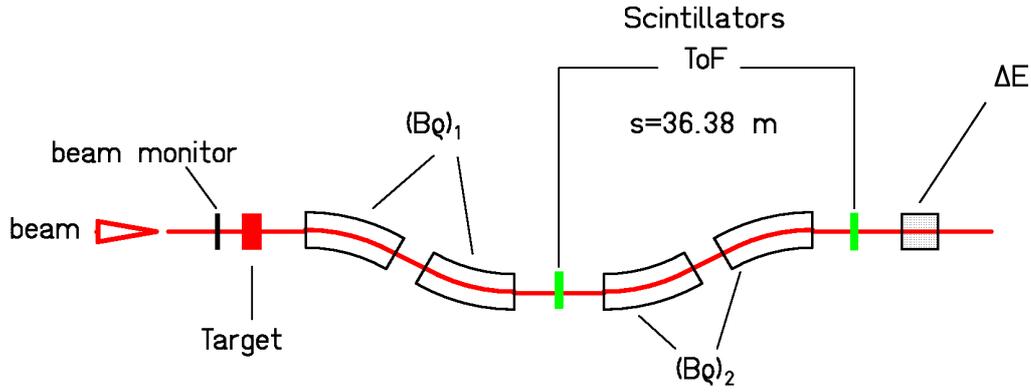

**Figure 1**: Schematic drawing of the fragment separator with the essential detector equipment, used in the experiment.

## 2.2. Isotopic identification

Having measured the magnetic rigidity, $B\rho$, the time-of-flight, and the nuclear charge, $Z$, of the ion, the mass number, $A$, of any reaction product could be determined according to the equation:

$$A = \frac{Z \cdot e}{m_0} \frac{B\rho}{c\beta\gamma} \qquad (1)$$

where $\beta \cdot c = v$ is the velocity of the ion, obtained from the time-of-flight, $\gamma$ is the relativistic parameter, $m_0$ is the nuclear mass unit, and $e$ is the electron charge. The full identification was possible for every nuclide produced, thanks to high resolutions in $Z$ and $A$ ($Z/\Delta Z = 140$, $A/\Delta A = 450$).

Once the nuclide was identified, its velocity could easily be determined by means of the equation:

$$\gamma \cdot v = B\rho \cdot \frac{Z \cdot e}{A \cdot m_0} \qquad (2)$$

with much higher precision that depended only on $B\rho$, since the mass $A$ and the charge $Z$ in equation (2) are integer numbers. This method gave an absolute measurement of the velocity and produced a very accurate result. The resolution in magnetic rigidity was about $3 \cdot 10^{-4}$. In case of many events, the mean value was even better defined.

## 2.3. Acceptance of the spectrometer

The high ion-optical resolution of the fragment separator is absolutely necessary in order to obtain full identification of the reaction products in mass and atomic number. This is the prerequisite for applying the isospin thermometer to the fragmentation reaction, which is the basic idea of the present work. Another property, which is important for the interpretation of the results, is the acceptance of the separator in angle and momentum. The acceptance of the separator is adapted to the emittance of heavy fragmentation products, but it does not fully cover the very light products. In the present chapter we will discuss, how the limited acceptance might lead to incomplete and eventually biased information.

### *2.3.1. Detection efficiency for the heavy residues due to limited acceptance.*

As already mentioned, the acceptance of the fragment separator is limited to about ± 1.5 % in momentum and 15 mrad in angle around the central trajectory [11]. In the reaction $^{238}$U + Pb at $1 \cdot A$ GeV, considered here, the measured standard deviation of the relative longitudinal momentum in the range covered by the experiment [8] varies from 0.4 % to 1.4 % for fragmentation products between $A = 175$ and $A = 75$. Since the angular distribution of the fragmentation products was not measured directly, we estimate it from the longitudinal momentum distribution by assuming that the velocity distribution of a certain nuclide is isotropic around its average velocity. According to results obtained with a full-acceptance set-up at the ALADIN spectrometer [12], this assumption seems to be well justified. This results in a standard deviation ranging from 2.1 mrad to 6.8 mrad over the same mass range from $A = 175$ to $A = 75$. Consequently, the angular range is covered by more than 90% for $A = 75$ and close to 100% for $A = 175$ by the angular acceptance of the fragment separator. Also the momentum distribution is covered to a great extent: If the magnetic fields are properly tuned for a specific nuclide, more than 70% is covered for $A = 75$, and more than 99 % is covered for $A = 175$ by the momentum acceptance of the fragment separator. In fact, the limited momentum acceptance of the fragment separator is less crucial, since the momentum distributions of all fragments was fully measured by superposing the results obtained with different settings of the magnetic fields [8].

We conclude that even for the lightest residues studied in the considered experiment [8], the bulk part of the distributions in momentum and angle is covered by the fragment separator and thus included in the measurement. A small part of at most 10% in the tails of the angular distribution, not covered by the angular acceptance of the FRS, is not expected to have any decisive influence on the observations described later.

Although the separator covers the major part of the production of a certain nuclide in a specific setting, the limited acceptance in magnetic rigidity does not allow to determine the multiplicity distribution of the reaction products in one event. The specific feature of the experiment [8] is the precise determination of nuclear charge, mass and velocity of the heavy fragment detected. This information is complementary to that obtained in large-acceptance devices. The distribution of elements simultaneously produced in the fragmentation of gold is known from previous work, e.g. performed at ALADIN [13]. From those results one can deduce that the products considered in the present work ($75 \leq A \leq 175$) are most likely the respective largest fragment emerging from each individual reaction.

*2.3.2. Missing information from nucleons and light clusters*

It is obvious that the fragment separator is not suited to give kinematically complete and exclusive information on the fragmentation reaction. In particular, very light clusters and nucleons, which populate a large range in angle and momentum, can only be covered to a small extend. In fact, light elements were not registered at all in the present experiment [8] due to the detector thresholds. Of course, it would be desirable to have a complete coverage and a full identification in mass and atomic number of all reaction products in a single experiment. However, with the experimental installations presently available, this goal cannot be achieved. Therefore, the rather complete information on all reaction products obtained with devices like INDRA or ALADIN with their rather limited nuclide identification for heavy residues should be combined with the information on the full nuclide identification of the heaviest residue emerging from the reaction as obtained with high-resolution spectrometers like the fragment separator. We hope that the present work gives the impetus for a combined analysis of all information accessible by the different experimental approaches in order establish a picture of the reaction mechanism as complete and as precise as possible. To reach this aim in the same experiment, a new generation of spectrometer which combines high resolution with high detection efficiency is required.

## 2.4. Discrimination of fission events

The experiment [8] allowed to identify those residues which were produced by fission by means of their velocity distribution. As an example, figure 2 shows the longitudinal velocity distribution in the beam frame for $^{94}$Zr. For this neutron-rich isotope, the fission contribution is rather strong. The distribution presents a central hump, centred approximately on the beam velocity, and two external humps, corresponding to large absolute values of the velocity. These triple-humped distributions can be interpreted taking into account the limited angular acceptance of the FRS: Only those fragments with a small perpendicular velocity component are transmitted. The central hump results from fragmentation reactions, while the external humps result from fission. Fission fragments emitted at backward direction are centred at –1.3 cm/ns, those emitted in forward direction are centred at +0.9 cm/ns, while the fission fragments emitted in sideward direction are suppressed. The fact that the mean velocities of both, fission fragments and fragmentation products, are slightly negative originates from the longitudinal momentum transfer to the respective projectile spectator in the abrasion process.

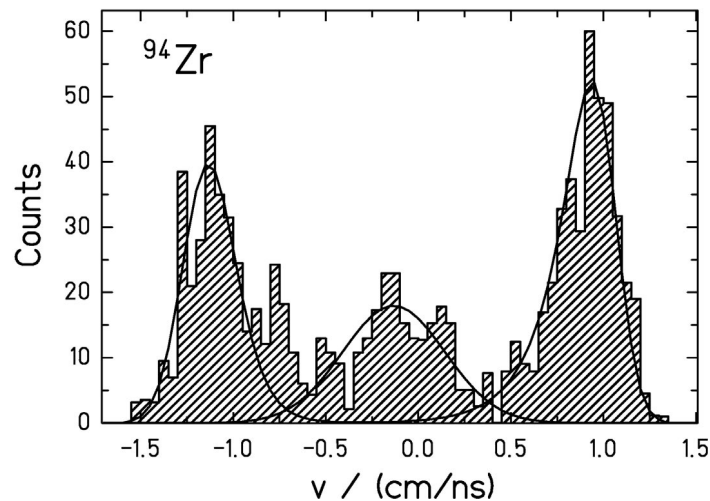

**Figure 2:** Measured velocity distribution of $^{94}$Zr produced in the reaction $^{238}$U + Pb at 1·$A$ GeV (data from Reference [8]). The velocities are given in the frame of the beam in the middle of the target. The three curves show the deduced contributions from fission backward, fragmentation, and fission forward, respectively.

The interpretation of the external humps as fission events was confirmed by the theoretical expectations for the velocities of the fission products [14], which gave values consistent with the experimental results. This aspect is intensively discussed in Reference [8]. The selection of non-fission products can easily be done by filtering only those events that fill the central hump of the velocity distribution. The velocity pattern, demonstrated in figure 2 for the nuclide $^{94}$Zr, is an unambiguous signature for distinguishing the two reaction mechanisms, which was applied in many experiments [15, 16, 17, 18, 19, 20, 21, 22, 23, 24].

It has been proposed [25] that fission-like events may also originate from a binary decay in a break-up process from a low-density system due to surface instabilities. One might expect that these products are characterised by lower kinetic energies, due to the large break-up volume and by lower *N*-over-*Z* ratios due to the higher initial excitation energies if compared to conventional fission products. The velocities of the nuclei observed in the outer velocity humps are rather consistent with those expected for fission products that originate from the fission competition in the statistical de-excitation cascade. Also the fact that the fission products are on the average much more neutron-rich than the fragmentation products of the same element indicates that they predominantly originate from moderately excited fissile pre-fragments. That means that direct evidence for the proposed binary break-up process was not found. Therefore, we consider that the fission fragments do not carry any information on the break-up of highly excited systems, and consequently we discard them from the present analysis.

## 2.5. Results

Using the values[**] of the formation cross sections of the nuclides produced in the fragmentation processes [8], it was possible to reconstruct the isotopic distributions for every observed element and to calculate the mean neutron-to-proton ratio. In figure 3, the measured cross sections of zirconium isotopes are shown.

In figure 4, the so-obtained values of the mean *N*-over-*Z* ratio of the fragmentation products for four different projectile-fragmentation reactions are presented as a function of the nuclear charge, *Z* (see figure caption for the details). All the presented experimental data were measured with high-resolution magnetic spectrometers. Fission events are never included. The experimental results are compared with two "reference" lines. The solid line represents the stability valley; it was calculated using the following formula, taken from Reference [26]:

$$Z = \frac{A}{1.98 + 0.0155 \cdot A^{2/3}} \qquad (3)$$

---

[**] The numerical values are available on the WEB site http://www-wnt.gsi.de/kschmidt/EnB99.htm

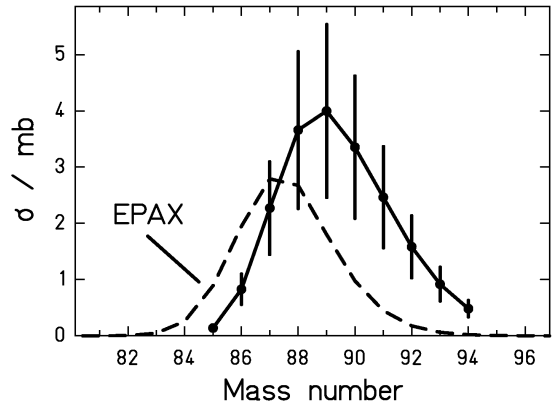

**Figure 3:** Measured cross sections of zirconium isotopes produced by fragmentation in the reaction $^{238}$U + Pb at 1·$A$ GeV (data points) compared to the prediction of the EPAX code [27] (dashed line). The error bars include systematic uncertainties.

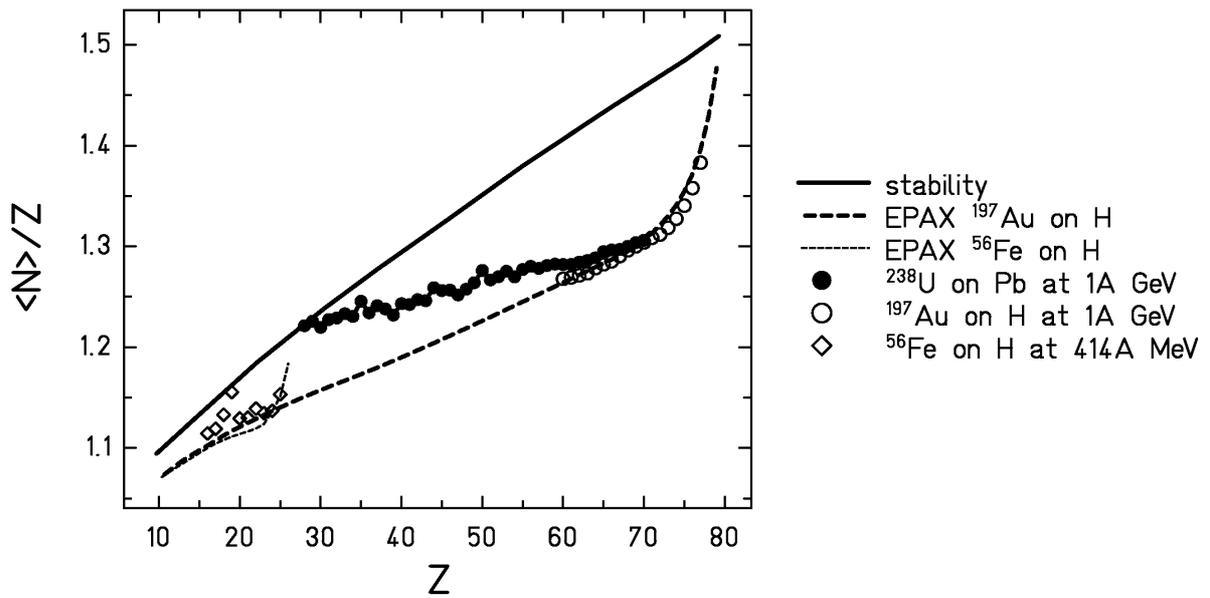

**Figure 4:** Comparison of the mean neutron-to-proton ratio found in the isotopic distributions after fragmentation of $^{238}$U in lead at 1·$A$ GeV [8] with the beta-stability line [26] and the prediction of EPAX [27] for the fragmentation of $^{197}$Au and for the fragmentation of $^{56}$Fe. The measured mean values from the fragmentation of $^{197}$Au in hydrogen at 800·$A$ MeV [20] and of $^{56}$Fe in hydrogen at 414·$A$ MeV [28] are shown in addition. Fission events are excluded.

The dashed line is the result of a calculation for $^{197}$Au on H$_2$ performed with EPAX [27], a semi-empirical code based on the idea that fragmentation products result from long, sequential evaporation chains, at the end of which the proton-evaporation and the neutron-evaporation probabilities are approximately the same. In this scenario, which represents one of the characteristics of the limiting fragmentation, it is expected that the fragmentation products generated with such a mechanism eventually land on a universal and rather narrow corri-dor on the chart of the nuclides. For the reactions $^{197}$Au on H$_2$ at 800·$A$ MeV and $^{56}$Fe on H$_2$ at 414·$A$ MeV the accordance with EPAX is excellent, and the sequential evaporation picture seems to be the valid one. Since the higher the excitation energy of the pre-fragment is, the more valid the hypothesis of limiting fragmentation should be, we would expect to find an even more perfect overlap between the mean $N$-over-$Z$ ratio of the fragmentation products for $^{238}$U on Pb and the EPAX prediction. Surprisingly enough, the experimental data show larger mean values, with a tendency to even cross the stability line for products with nuclear charge below ≈ 28. Compare also figure 3, where the full isotopic distribution for Z = 40 is shown together with the EPAX prediction.

Indications for deviations of the neutron-to-proton ratio of fragmentation products from EPAX were already reported by Lindenstruth [12] for the reactions $^{197}$Au on copper and on aluminium. As we will see in section 3.2.2, gold is not the best choice as projectile beam for the application of the isospin thermometer due to its relatively low neutron excess. Therefore we did not exploit Lindenstruth´s data for our purpose.

## 3. Discussion

### 3.1 Simultaneous break-up versus sequential decay

In the present context, it is useful to consider the evaporation stage of an excited nucleus as a diffusion process on the chart of the nuclides with elementary steps characterised by the evaporation of neutrons, protons and light nuclei. This idea was introduced by Campi and Hüfner [29] and later refined by Gaimard and Schmidt [30]. A basic feature, which is of prime importance, is that the diffusion process tends to enter into a universal evaporation corridor [31], where the evaporation probabilities of essentially protons and neutrons reach their asymptotic values. Later, the evaporation corridor was also denoted as evaporation attractor line [32]. The evaporation corridor is situated between the beta-stability line and the proton drip line. Only if the excitation energy at the beginning of the evaporation cascade is not sufficiently high, the residues keep some memory of the $N$-over-$Z$ ratio of the initial excited nucleus, but with every evaporation step, this memory is more and more lost. The universal behaviour of the $N$-over-$Z$ ratio of evaporation residues is the prerequisite for semi-empirical parameterisations of the cross sections of heavy residues from fragmentation reactions like EPAX [27]. In this parameterisation, it is assumed that the lighter fragments are formed from more violent collisions, leading to longer evaporation chains. That means, from a certain mass loss on, the shape of the isobaric distributions is universal. It neither depends on the projectile nor on the target nucleus. This assumption seems to be confirmed by a great variety of data. Among these data, in figure 4 only two systems are presented, $^{197}$Au on protons and $^{56}$Fe on protons.

The data of the reaction $^{238}$U on lead shown in figure 4 show a different behaviour. Residues below $Z = 70$ tend to deviate from the evaporation-residue corridor more and more. Lighter residues obviously experienced an evaporation cascade which was too short to reach the evaporation corridor. *More violent collisions seem to introduce lower excitation energies.* This is a quite surprising finding which demands for a careful analysis.

An essential feature that is necessary in order to reach the asymptotic values of neutron and proton evaporation is the sequential character of the evaporation process. The result of the preceding step is the initial condition of the next one. A simultaneous break-up does not show this feature. It has two essential characteristics: Firstly, the excitation energy is shared among the clusters and, secondly, the *N*-over-*Z* ratio of these clusters is expected to be close to the one of the total decaying system [33]. It is just slightly modified for individual fragments by charge polarisation or isospin fractionation. It seems to be worthwhile to test the hypothesis that the lack of apparent excitation energy found in the fragmentation products of $^{238}$U is related to a break-up stage before the evaporation cascade. Such a break-up stage is known to occur at high excitation energy, leading to multifragmentation, i.e. the formation of intermediate-mass and light fragments. The neutron excess of the heavy fragmentation products might indicate another manifestation of nuclear instabilities at relatively small excitation energies, similar to multifragmentation. The basic process behind could be the nuclear liquid-gas phase transition.

We would like to mention that the systematic shift in neutron-to-proton ratio with respect to the evaporation corridor, marked by the EPAX parameterisation, as discussed above has not been realised before, neither in the data from the high-resolution spectrometer FRS [8] nor in average values measured with ALADIN [34]. The deviations of the measured isotopic distributions from the evaporation corridor became more evident, since the new parameterisation of EPAX [27] became available which gives a more reliable parameterisation of the evaporation corridor on the basis of recent experimental data (e.g. [16, 18]), in particular for heavy nuclei.

In the following, we will introduce the isospin thermometer and apply it to determine the temperature at the beginning of the evaporation stage. After some schematic considerations, we will use a three-stage reaction model and the statistical multifragmentation model to calibrate the isospin thermometer. Our aim is to come to a consistent description of the heavy-residue production in relativistic heavy-ion collisions.

### 3.2 Application of the isospin-thermometer method

*3.2.1 Principle of the isospin thermometer*

The isospin thermometer is a specific method to deduce the temperature of nuclear systems from the isotopic distributions of the residues at the end of the evaporation cascade. The method consists of applying an evaporation code with the quite well known ingredients of the statistical model in order to deduce the temperature at the beginning of the evaporation cascade. It has been applied in Reference [15] to deduce the mean excitation energy induced in peripheral relativistic heavy-ion collisions as a function of mass loss. In this way, a value of 27 MeV per abraded nucleon has been deduced.

*3.2.2. Basic idea to determine the freeze-out temperature*

Figure 5 gives a schematic interpretation of the findings reported here. *In this sub-section, we do not intend to give a quantitative answer, but rather to illustrate the basic idea of our method.* In these considerations we neglect the influence of fission in all aspects. Several additional approximations are made in this sub-section in order to work out the basic concept in the clearest way. For a quantitative analysis, we refer to the calculations with more elaborate models, presented in section 3.3 and 3.4.

We assume that the break-up produces clusters of different sizes with the *N*-over-*Z* ratio of the $^{238}$U projectile. They all lie on the dotted line. Their excitation energy leads to an evaporation cascade. Since the primary clusters are very neutron-rich, neutron evaporation dominates, and depending on the initial excitation energy, the products loose part of their neutron excess. If the excitation energy is sufficiently high, they enter into the evaporation corridor, where the competition between neutron and proton evaporation reaches its asymptotic value. The evaporation-residue corridor, as predicted by EPAX, is indicated by the dashed line. Further on, we assume that all clusters are formed with the same freeze-out temperature. This leads to excitation energies of

$$E_{freeze\text{-}out} = a\,T_{freeze\text{-}out}^{2}. \tag{4}$$

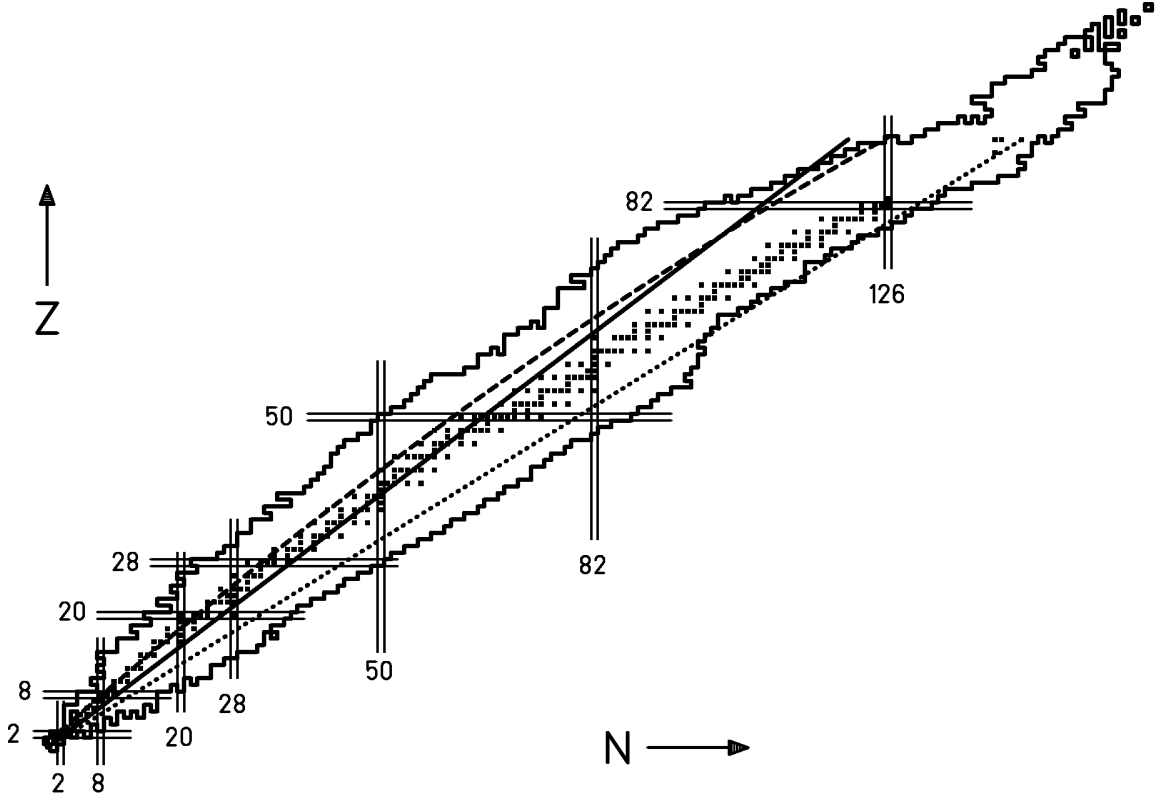

**Figure 5:** Schematic illustration of the application of the isosopin thermometer to determine the temperature at freeze-out of the break-up configuration in the fragmentation of $^{238}$U. Dotted line: Mean *N/Z* of the primary clusters after break-up. Dashed line: Evaporation-residue corridor. Solid line: Position of the final clusters after the evaporation cascade if only neutron evaporation is assumed. Full squares mark the primordial nuclides. The full surrounding lines indicate the limits of known nuclides. See text for more details.

With the level-density parameter set to $a = A/10$, and assuming a guess value of $T_{freeze\text{-}out} = 4$ MeV, we get $E_{freze\text{-}out} = 1.6 \cdot A$ MeV. We can reasonably assume that from the beginning of the evaporation process only neutrons are evaporated, until the evaporation of neutrons and protons reach their asymptotic values. We further assume that 15 MeV are needed to evaporate one neutron. This value results from estimating the expression $S_n + 2T$ (neutron separation energy plus kinetic energy of the neutron which amounts to two times the nuclear temperature on the average). At the beginning of the cascade, the binding energy of the neutron is low, the average kinetic energy it acquires is high. At the end of the cascade, the situation is reversed, with the net result that the sum of the binding and kinetic energies is approximately constant and equal to 15 MeV. Under these conditions, all end-products have the same *N*-over-*Z* ratio of

$$N_{final} / Z_{final} = (N_{initial} - 1.6\, A_{initial}/15) / Z_{initial} \approx 1.3, \qquad (5)$$

corresponding to the solid line in figure 5. The solid line marks nuclei which are more neutron-rich than the evaporation-residue corridor below $Z \approx 70$, and even more neutron-rich than the beta-stability line below $Z \approx 40$.

These characteristics are in remarkable qualitative agreement with the findings reported in the preceding chapter. Thus, it seems that the hypothesis of a simultaneous break-up process which precedes the statistical evaporation is compatible with the behaviour of the mean neutron-to-proton ratio of the fragmentation observed. Attempts for a more quantitative interpretation will follow in the following chapters.

It is clear that the restriction to neutron evaporation is unrealistic for elements above $Z \approx 70$ where the final fragments according to the solid line would be more neutron deficient than the evaporation-residue corridor. In this range, the *N*-over-*Z* ratio is expected to saturate at the evaporation corridor. Only for $Z < 70$ the excitation energy of the primary clusters is not sufficient to reach the evaporation-residue corridor at the end of the evaporation cascade. They preserve part of the large neutron excess of the $^{238}$U nucleus. This allows applying the isospin thermometer.

From these considerations it becomes clear that the method applied in the present work to determine the freeze-out temperature with the isospin thermometer works best in the fragmentation of a very neutron-rich nucleus like $^{238}$U. Assuming an unchanged freeze-out temperature, the fragmentation of lighter, less neutron-rich nuclei, would also produce less neutron-rich primary clusters. Consequently, the evaporation process would lead to final fragments closer to, or eventually even on the evaporation-residue corridor, where the isospin thermometer would saturate. Therefore, $^{238}$U was the best choice as a projectile, although the strong fission competition introduces some complications. However, this was not a severe problem, since fission products could safely be suppressed due to the high-resolution detection device.

From this argumentation, we may also deduce that the *N*-over-*Z* ratio of fragmentation-evaporation residues, even for masses appreciably lighter than the projectile, is not universal. Due to the limited freeze-out temperature, they are influenced by the *N*-over-*Z* ratio of the initial nucleus which is fragmented. That means that the concept of limiting fragmentation, which is behind some semi-empirical parameterisation of residue cross sections (e.g. EPAX), should be considered with caution.

### 3.3 Comparison with a three-stage model

We have used a three-stage model with an abrasion stage [30], a break-up stage and an evaporation stage [18] for a more detailed interpretation of the available data. In the abrasion stage the projectile spectator acquires an excitation energy of 27 MeV per nucleon removed [15]. If this energy leads to temperatures above the freeze-out temperature, we assume that the system undergoes a break-up stage. At this stage, part of the initial energy is removed through the loss of mass in form of nucleons or light clusters, which is not specified explicitly. Best agreement with the experimental data is achieved if the energy consumed to lose one mass unit varies from 8 MeV for an initial temperature of 5.5 MeV to 4 MeV for an initial temperature of 10 MeV; the higher the temperature, the smaller the consumed energy per unit mass loss is. At first glance, this tendency is opposite to the expectation that the particles should be emitted with higher kinetic energies when the temperature of the source increases. The explanation could be that the average size of the clusters emitted increases with increasing temperature, since the remaining binding energies of the clusters would tend to reduce the energy release in the break-up process. Please note that the value of this energy reduction only enters into the absolute cross sections, not into the shape of the final isotopic distributions. Note also that the multiplicity of the emitted clusters does not affect the evaluation of the cross section of large residual fragments, since the probability that two of them are generated in the same (non-fission) event is zero if the mass of the break-up product exceeds half the mass of the projectile and it is still negligible over the whole mass range of the experiment considered according to References [12, 13, 30]. We assume that in the break-up stage the *N*-over-*Z* ratio is conserved, so the break-up product has the same *N*-over-*Z* ratio as the projectile spectator after abrasion. Finally, a standard evaporation [18] follows for every pre-fragment. If the temperature of the system after the abrasion is lower than the freeze-out temperature, the spectator is considered to be the only pre-fragment and the evaporation starts immediately.

The results of such a calculation with different values assumed for the freeze-out temperature are shown in figure 6. A remarkable agreement with the data is found, if a value of 5 to 5.5 MeV is used for the freeze-out temperature in all cases. A good agreement is obtained also in the evaluation of the cross sections, as shown in figure 7, where the results of the three-stage model are compared with the experimental data for the isotopic distributions of four elements. The figure includes also the predictions of the statistical multifragmentation model, which will be discussed in the next section. The widths of the isotopic distributions predicted by the three-stage model are basically consistent with the experimental ones, they seem to be even a bit overestimated by the calculations. Note that the calculations were performed assuming a sharp value of the freeze-out temperature of 5 MeV. This fact lets little room for introducing a fluctuation of the freeze-out temperature in the calculation, because this would even increase the width of the isotopic distributions. Fluctuations in the *N*-over-*Z* ratio of the prefragments after abrasion and in the kinetic energies of the evaporated particles seem to be sufficient to explain the observed width in the isotopic distributions.

**3.4 Comparison with SMM calculations**

The method applied in the preceding section to back-trace the conditions at the beginning of the evaporation cascade from the mean values of the observed isotopic distributions was based on the assumption that all products of the break-up stage have the same *N*-over-*Z* ratio. This assumption is not made in the statistical multifragmentation model [35, 36, 37]. Within this model, a microcanonical ensemble of all break-up channels composed of nucleons and excited fragments of different masses is considered. It is assumed that an excited nucleus expands to a certain volume and then breaks up into nucleons and hot fragments. All possible break-up channels are considered. It is also assumed that at the break-up time the nucleus is in thermal equilibrium characterised by the channel temperature *T*. The probability $W_j$ of a decay channel *j* is proportional to its statistical weight:

$$W_j \propto \exp S_j(E^*) \, , \tag{6}$$

where $S_j$ is the entropy of the system in a state corresponding to the decay channel $j$, which depends on excitation energy $E^*$ and other parameters of the system. The fragments with mass number $A_f > 4$ are treated as heated nuclear liquid drops. Each fragment contributes the bulk-, surface-, Coulomb- and translational terms to the free energy and to the entropy of the system. The light fragments with $A_f \leq 4$ are considered as elementary particles having only translational degrees of freedom. The Coulomb interaction between all fragments and gas particles is taken into account via the Wigner-Seitz approximation.

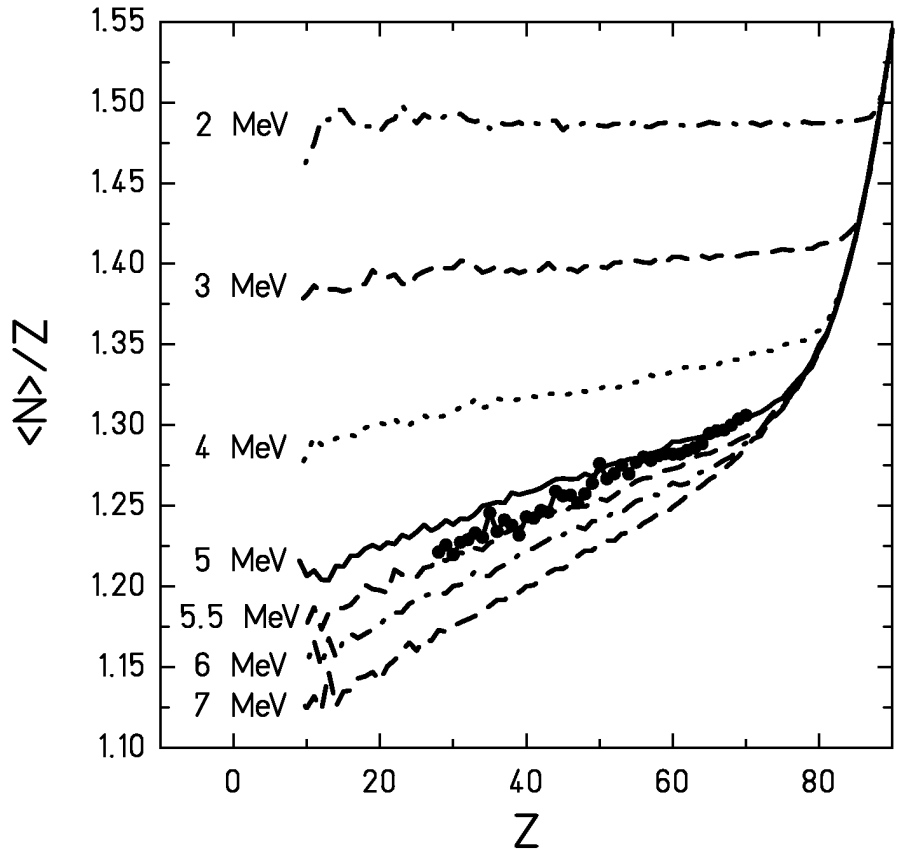

**Figure 6:** Experimental data on mean neutron-over-proton ratios of heavy fragmentation-evaporation residues produced in the fragmentation of $^{238}$U + Pb (data points) in comparison with the results of a three-step nuclear-reaction model. It considers abrasion, break up and sequential decay (see text), using different values of the freeze-out temperature of the break-up stage.

The break-up channels are simulated by the Monte-Carlo method according to their statistical weights. In the present calculations it was assumed that the break-up occurs at a density of one sixth of the normal nuclear density. After break-up of the system, the fragments propagate independently in their mutual Coulomb fields and undergo secondary decays. The de-excitation of large fragments ($A_f > 16$) is described by the evaporation-fission model, and for smaller fragments by the Fermi break-up model [36, 1]. Presently, the SMM is one of the most successful models to describe fragmentation and multifragmentation data (see e.g. References [1, 4, 34, 37, 38]).

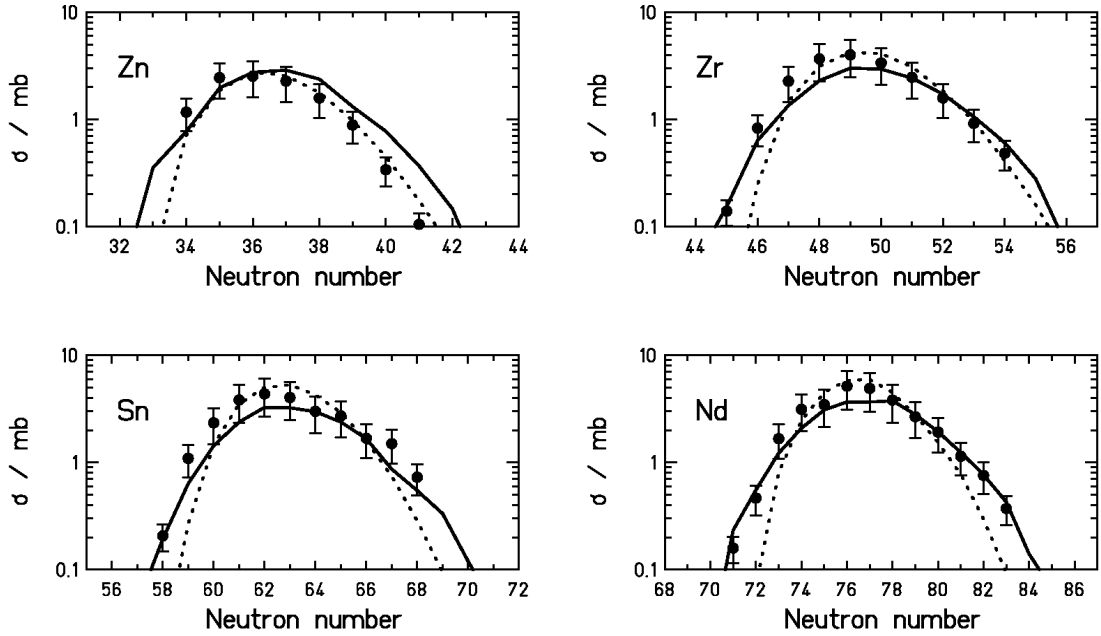

**Figure 7:** Measured cross sections of zinc (Z=30), zirconium (Z=40), tin (Z=50), and neodymium (Z=60) isotopes produced by fragmentation in the reaction $^{238}$U + Pb at 1·A GeV (data points) compared to the prediction of the three-stage code (full line) and to the prediction of the SMM code (dotted line). The SMM results were normalised to the sum of the measured cross sections of each element. In both model calculations, the freeze-out temperature of 5 MeV was used.

Since the heavy residues observed in the experiment indicate that rather peripheral collisions are involved, the approximation is made that the initial system at break-up still contains all nucleons of the projectile. The excitation energy at break-up was taken as a parameter of the calculations. This also defines the temperature of the system.

The mean *N*-over-*Z* ratio of the isotopic distributions that results from these calculations is compared to the experimental values of the reaction $^{238}$U on lead at 1 *A* GeV [8] in figure 8. The excitation energies of 1, 1.5, 2, 2.5 and 8 MeV/u correspond to mean temperatures of 3.1, 3.9, 4.4, 4.9 and 7 MeV, respectively. With increasing temperature, the mean neutron excess approaches the evaporation-residue corridor, marked by the EPAX prediction. In the beginning, this trend is fast, then it slows down until the temperature saturates in the region of the phase transition at the level of 5 to 6 MeV, and then it increases again in the gas phase. The corridor is finally reached at very high excitation energies.
There is a remarkable agreement found with the data when an energy around 2.5 MeV per nucleon is assumed. This corresponds to a freeze-out temperature around 5 MeV. Firstly, it is this value which has to be considered as the essential result of our analysis. Secondly, it is an important finding that the freeze-out temperature is independent of the mass of the residue we observe. The experiment does not sort the observations according to impact parameter. Therefore, the observed fragments of a given size emerge from different initial conditions. Nevertheless, it is a remarkable finding that collisions, which predominantly produce relatively light clusters proceed by the same freeze-out temperature as those which predominantly produce very heavy clusters. This seems to be an indication that the freeze-out temperature is rather independent of the initial conditions. Obviously, the more sophisticated treatment of the *N*-over-*Z* ratio of the products of the break-up process in the SMM model leads to an even better agreement with a constant value of the freeze-out temperature than the three-stage model described above.

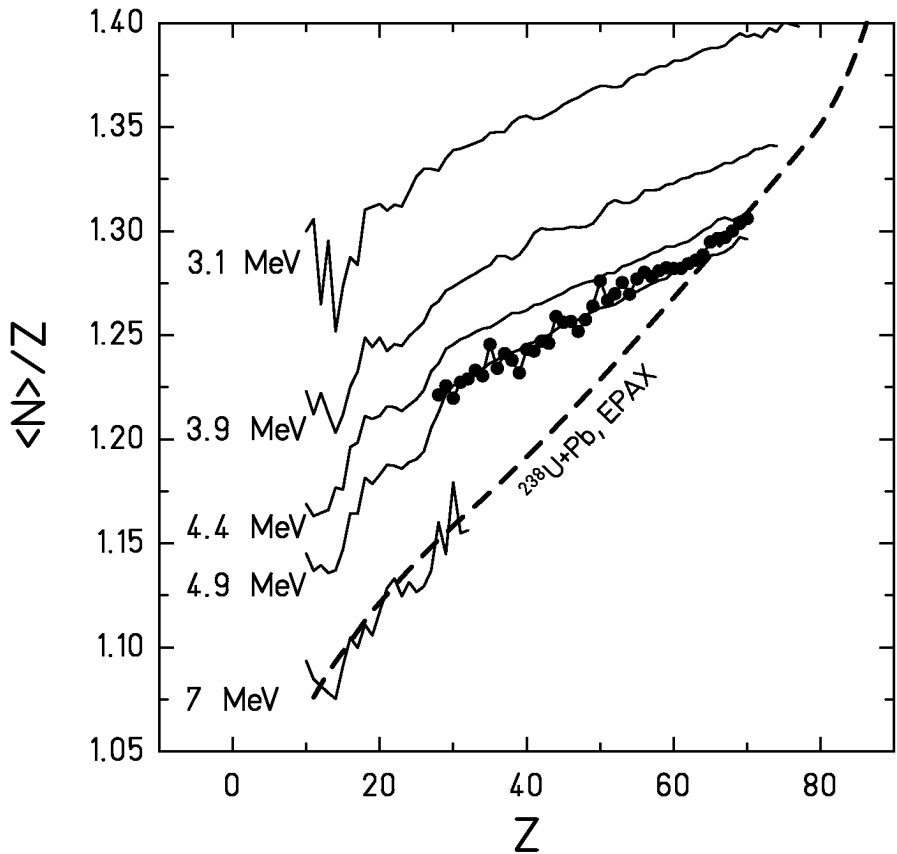

**Figure 8:** Mean neutron excess of the isotopic distributions of heavy residues obtained with the SMM model with different values of the temperature in comparison with the experimental values of the reaction $^{238}$U on lead at 1·$A$ GeV [8] (full points). The prediction of EPAX [27] for the fragmentation of $^{238}$U (dashed line) is shown in addition.

We investigated, to which extend the input conditions of the SMM calculation affect the validity of this result. In the previous calculations a value of the density 6 times smaller than the normal nuclear density was chosen. When the calculations are performed with a density 3 times smaller than the normal nuclear density, the results agree with the $^{238}$U+Pb data with a mean value of the freeze-out temperature of 5.1 MeV instead of 4.9 MeV. Also a variation of the mass of the system at break-up is not expected to have an essential influence on the $N$-over-$Z$ ratio of the heavy residues in the calculation, as long as the $N$-over-$Z$ ratio of the initial system is not changed. For this reason, we consider the value of 5 MeV substantially correct.

In figure 7, the SMM isotopic distributions were normalised to the experimental data. In fact, the SMM calculation was performed starting always with a fixed charge and mass of the system (Z=92, A=238). That means that the mass loss in the abrasion phase is neglected, which is anyhow small for the relatively heavy residues observed according to the three-stage model discussed above. In this way, the calculation is representative for very peripheral collisions only. It is intended to yield the shape of the isotopic distributions, not a quantitative prediction of the cross sections. As it was shown in [1, 34, 37], to reach a quantitative description, the SMM should be combined with preequilibrium models. The SMM calculation reproduces the shapes of the isotopic distributions rather well, although they tend to be too narrow for the heavier elements. This could be explained by the fact that this SMM calculation did not consider any fluctuation in the N-over-Z ratio of the prefragments, while this effect was included in the abrasion process of the three-stage model. It is known that fluctuations of the temperature tend to broaden the isotopic distributions. The fact that the measured widths are relatively small give us confidence that there are no substantial fluctuations of the freeze-out temperature. Within the SMM model, the variance of the temperature around the mean value is less than 0.5 MeV. Changes in the input values of the density in the input of the SMM calculation affect only the height of the cross sections.

**3.5 General understanding of heavy-residue production in relativistic heavy-ion collisions**

When combining the conclusions drawn from the application of the isospin thermometer with other experimental information available, we come to a rather general understanding of heavy-residue production in relativistic heavy-ion collisions. Figure 9 shows a summary of the results obtained with the isospin thermometer. The excitation energy at the beginning of the evaporation cascade is shown as a function of the atomic number of the pre-fragment, entering into the evaporation cascade. Since neutron evaporation dominates, the picture would not look much different, if drawn as a function of the atomic number of the *final* fragment.

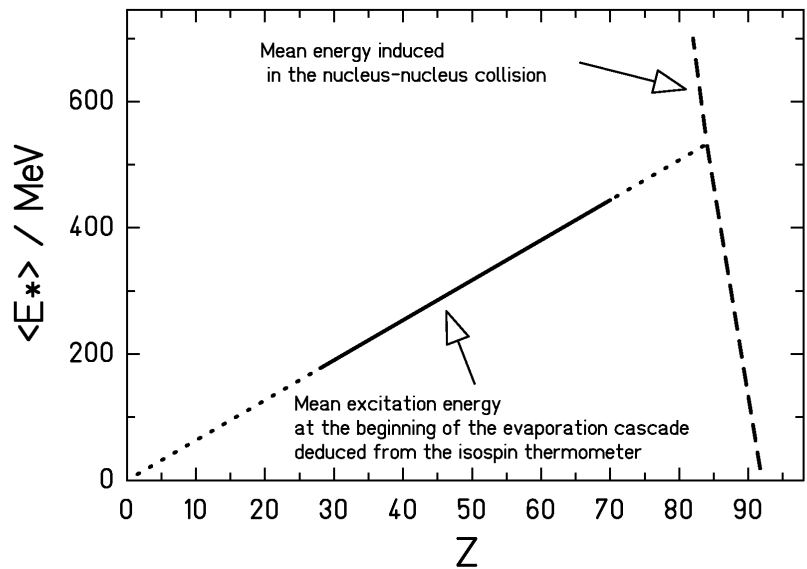

**Figure 9:** Schematic presentation of the initial energy induced in the abrasion stage and of the mean excitation energy at the beginning of the evaporation cascade after fragmentation of $^{238}$U in a lead target as deduced with the isospin-thermometer method. Close to the projectile, the excitation energy induced increases linearly with the mass loss, determined in Reference. [15] (dashed line). At $Z \approx 84$, the temperature of 5 MeV is reached in the abrasion, and the break-up sets in. The energy introduced into the evaporation cascade linearly decreases with decreasing mass number, corresponding to the constant value of 5 MeV of the freeze-out temperature (dotted line). The range covered by the present work is marked by a full line.

When starting from very peripheral collisions, few nucleons are removed from the projectile, leaving an excited heavy spectator nucleus. According to the results obtained in the fragmentation of $^{197}$Au in an aluminium target [15], the average excitation energy amounts to about 27 MeV per abraded nucleon. We assume that the result of the fragmentation of gold in aluminium can directly be applied to the fragmentation of uranium in lead. This seems to be justified from the different experimental results on several systems, which have been accumulated in several experiments [16, 18]. The difference in binding energy of protons and neutrons should not have any influence on the abrasion phase. Therefore, protons and neutrons are expected to be abraded according to their abundance in the projectile. That means that one proton is accompanied by 1.59 neutrons on the average in the fragmentation of $^{238}$U. This leads to an excitation energy of about 70 MeV introduced per proton abraded.

It should be noted that the isospin thermometer cannot directly be applied to the fragmentation of $^{238}$U in this near-projectile region, because the isotopic distributions are severely modified by the fission competition [18]. However, fission is not expected to have any influence on the validity of figure 9 concerning the excitation energy induced, since the fission competition just adds up to the possible decay channels of the excited equilibrated system.

Near $Z = 84$, the excitation energy induced corresponds to a temperature of 5 MeV. It is to be expected that the excitation energy induced in the nuclear collision increases further for an increasing number of protons abraded, but the high values of the dashed line are not found in the evaporation cascade of the final fragments, because the system rather looses part of its energy first by a kind of break up. The evidence for the real nature of these reactions has to come from a combined analysis, including the information from kinematically more complete measurements. However, it is the remarkable result of the present analysis that there is a limiting temperature of 5 MeV of any equilibrated system which enters into the statistical de-excitation cascade. We emphasize that this result is obtained by using two different kind of models: One is adjusted for the description of spallation processes, another one for fragmentation and multifragmentation processes. Fission processes are taken into account in both cases. The experimental signature from the data of Reference [8] directly relates to the range between $Z = 28$ and $Z = 70$, marked by the full line in figure 9.

From the statistical multifragmentation model, no direct conclusion on the dynamical evolution of the system before forming a thermalised nuclear system can be made. Together with the rich information gathered on multifragmentation from the observation of light clusters, however, one may speculate that the nucleus expands after the abrasion process due to thermal pressure. As a result of density fluctuations during the expansion, light and intermediate-mass fragments are produced firstly. In this way, the nucleus looses its excessive energy and a large residue survives with a temperature around 5 MeV.

While, in accordance with the abrasion model, there is a simple geometrical relationship expected between the atomic number of the heaviest fragment and the impact parameter above $Z = 84$, this relationship is less direct below $Z = 84$. Here, the charge found in the heaviest initial cluster after the break-up does not comprise all protons of the heated spectator nucleus. Part is rather lost by the simultaneous formation of light clusters and single protons during break-up. There is still a correlation between the impact parameter and the charge of the heaviest fragment, as has been discussed in previous work of the ALADIN and EOS groups [34, 38, 39]. However, the simple correspondence deduced from the abrasion model is not correct.

A very similar conclusion as drawn in the present work has also been deduced from light charged particles emitted from intermediate-mass fragments in the range from $Z = 2$ to $Z = 30$ produced in central collisions of $^{129}$Xe + $^{nat}$Sn at 50 A MeV [40]: In this work with the fragment-particle correlation method, a limiting temperature around 5 MeV was deduced, below which "primary fragments deexcite only by evaporation". Thus, heavy-ion collisions in very different energy regimes show signatures of the same phenomenon.

It is very interesting to note that the mean limiting temperature deduced in the present work coincides with the kind of saturation value of about 5 MeV of the apparent temperature deduced from double isotopic ratios and the population of excited states in light clusters [2]. For example, as shown in [41], the temperature deduced from the IMF yields remains nearly constant at T ≈ 5 MeV in the very large excitation-energy range from 2.5 to 10 MeV/nucleon. Thus, the present analysis provides an important confirmation for features of the caloric curve deduced previously. Temperatures deduced from slopes of kinetic-energy spectra [2], which extend to much higher values, could essentially be influenced by pre-equilibrium emission and flow, and might not represent the values at freeze-out.

### 3.6 Consequences for the understanding of other phenomena in highly excited nuclei

In the experiment discussed, no evidence for the decay of heavy nuclei with mean temperatures exceeding 5 MeV has been found, although the centre-of-mass energy of the colliding nuclei exceeds the total binding energy of the system by about two orders of magnitude. The result of our analysis might indicate that the probability of a compound nucleus surviving, equilibrated in both intrinsic and collective degrees of freedom, drops fast above a limiting temperature of 5 MeV. This finding would have far-reaching consequences for the understanding of many phenomena in highly excited nuclei.

We would like to comment on the consequences of the present discussion on the understanding of nuclear dissipation in fission from dedicated experiments. Signatures of dissipation are often deduced from the absence of fission occurring at high excitation energies. It is obvious that the finding of a limiting temperature of 5 MeV presented above introduces a limitation to fission as to any other statistical decay channel, which has no direct connection with nuclear dissipation. Therefore, the analysis of some corresponding experimental data should be revisited in the light of the present findings. A deeper discussion of this subject is found in Reference [42].

In another research field, the question of barriers for light charged-particle emission from highly excited nuclei, a long-standing controversy might be connected with the present findings. When a composite system is produced with sufficient energy induced, the nucleus might expand to a larger volume. This could explain the lowering respectively the washing out of the Coulomb barrier for charged particles emitted at the beginning of the de-excitation process, before a compound nucleus is formed (see also Reference [1]).

There are certainly many more observations, which might be connected with the possible non-existence of a fully equilibrated compound nucleus above the limiting temperature. A more careful discussion of this subject is beyond the scope of the present publication.

## 3.7 Outlook

It has become evident that the isotopic distributions of heavy residues contain important information on the break-up of highly excited nuclear systems, which might be linked to the nuclear liquid-gas phase transition. The data from the experiments analysed in the present work only allow a first glance on the topic, further investigations, which extend the knowledge in several directions are highly desirable. Some of the aspects to be investigated are the dependence of the experimental signatures of projectile break-up on the target size, on the beam energy and on the neutron excess of the projectile. Another important issue is to extend the investigations to lighter fragments where the difference in the neutron-to-proton ratio between the fragmentation products and the evaporation-residue corridor is expected to increase even further. This is also important to establish a closer link to the numerous investigations available on even lighter intermediate-mass fragments.

The final goal would be to study the characteristics of relativistic heavy-ion collisions with an advanced experimental set-up, which simultaneously allows obtaining a kinematically complete information on all nucleons, intermediate-mass fragments and heavy residues produced and identifying all these products unambiguously in mass and nuclear charge with high resolution. The design of such a set-up for GSI, based on a superconducting large-acceptance magnetic spectrometer is actually in progress [43].

## 4. Conclusion

In the fragmentation of $^{238}$U, residual nuclei below $Z = 70$ were found not to reach the evaporation-residue corridor, but to preserve part of the neutron excess of the projectile. Our analysis with two different models, the three-stage model and the SMM, gives indications for a change in the de-excitation mechanism with increasing excitation energy and the transition to a simultaneous-break-up phase. The *N*-over-*Z* ratio of the final fragment was used as an isospin-thermometer method in order to deduce the mean temperature at freeze-out to $T \approx 5$ MeV. This value was found to be independent of the size of the final fragment.

On the basis of the statistical interpretation, one might conclude that the probability for a compound nucleus surviving, equilibrated in both intrinsic and collective degrees of freedom, and entering into the evaporation cascade, should decrease considerably above a limiting temperature of 5 MeV. This would have far-reaching consequences for many phenomena found in highly excited nuclear systems including the fission process.

Looking at the phenomenon as a manifestation of the nuclear liquid-gas-type phase transition, the result represents new information on the freeze-out conditions, which confirms previous results based on temperatures deduced from isotopic ratios, the population of excited states of small clusters, and the correlation methods. It is consistent with a kind of saturation of the nuclear temperature as a function of excitation energy introduced over that range of excitation energy where heavy clusters are formed.

These results rely on the use of the most neutron-rich primordial nucleus available, on the full identification in *N* and *Z* of all heavy reaction products, and on the suppression of fission events by measuring the kinematical properties. These technical requirements were met by the heavy-ion accelerator SIS18 and the high-resolution spectrometer FRS at GSI Darmstadt.

## Acknowledgment

Fruitful discussions with P. Armbruster, W. Nörenberg, W. Reisdorf, and W. Trautmann, as well as with E. Plagnol and other members of the INDRA collaboration are gratefully acknowledged. This work has been supported by the European Union under contract nmbr. EC-HPRI-CT-1999-00001.